\documentclass[twocolumn,prl,amsmath,amssymb,showpacs,superscriptaddress,floatfix]{revtex4}
\usepackage{graphicx}% Include figure files
\usepackage{bm}% bold math

\usepackage{epsf}
\begin{document}
%\draft
\title{Phase diagram of the system with the repulsive shoulder potential
in two dimensions: density functional approach}%:
%First-Order versus Continuous Transition }
\author{E. S. Chumakov}
\affiliation{ Institute for High Pressure Physics RAS, 142190
Kaluzhskoe shosse, 14, Troitsk, Moscow, Russia}
\affiliation{Moscow Institute of Physics and Technology, 141700
Moscow, Russia}

\author{Yu.D. Fomin}
\affiliation{ Institute for High Pressure Physics RAS, 142190
Kaluzhskoe shosse, 14, Troitsk, Moscow, Russia}
\affiliation{Moscow Institute of Physics and Technology, 141700
Moscow, Russia}

\author{E. L. Shangina}
\affiliation{ Institute for High Pressure Physics RAS, 142190
Kaluzhskoe shosse, 14, Troitsk, Moscow, Russia}

\author{E. E. Tareyeva}
\affiliation{ Institute for High Pressure Physics RAS, 142190
Kaluzhskoe shosse, 14, Troitsk, Moscow, Russia}

\author{E.N. Tsiok}
\affiliation{ Institute for High Pressure Physics RAS, 142190
Kaluzhskoe shosse, 14, Troitsk, Moscow, Russia}

\author{V.N. Ryzhov}
\affiliation{ Institute for High Pressure Physics RAS, 142190
Kaluzhskoe shosse, 14, Troitsk, Moscow, Russia}
\affiliation{Moscow Institute of Physics and Technology, 141700
Moscow, Russia}

\date{\today}

\begin{abstract}
In the framework of the density functional theory of freezing
proposed in our previous works, we calculate the phase diagram of
two-dimensional system of particles interacting through the
repulsive shoulder potential. This potential consists of the hard
core and repulsive shoulder of the larger radius. It is shown that
at low densities the system melts through the continuous
transition in accordance with the
Kosterlitz-Thouless-Halperin-Nelson-Young (KTHNY) scenario, while
at high densities the conventional first order transition takes
place.
\end{abstract}


\maketitle

A large number of papers studying the melting transition in two
dimensions have been published during last decades. They include
results of real experiments, computer simulations and various
theoretical approaches. This is dictated by the growing interest
to the behavior of the nanoconfined systems. Confining drastically
changes the spatial distribution and the ways of dynamic
rearrangement of the molecules in the system. The confined fluids
microscopically relax and flow with characteristic times that
differ from the  bulk fluids. These effects play important role in
the thermodynamic behavior of the confined systems and can
considerably change the topology of the phase diagram. In general,
the motivation for the study of the confined systems follows from
the fact that there are a lot of real physical, chemical and
biological processes which drastically depend on the properties of
such systems \cite{rev1,rice,barbosa,buld2d,krott,krott1}.

It is not surprising that the spatial ordering of molecules
depends on the dimensionality of the space to which it is
confined. Mermin \cite{mermin} has shown that in in two dimensions
($2D$) the long-range crystalline order can not exist because of
the thermal fluctuations and transforms to the quasi-long-range
order. On the other hand, the real long range bond orientational
order does exist in this case. At high temperatures one can find
the conventional isotropic fluid.

The melting scenario in two dimensions is a subject of long
lasting controversy. Now it is widely believed that the
Kosterlitz, Thouless, Halperin, Nelson, and Young theory (KTHNY
theory) \cite{kosthoul73,halpnel1,halpnel2,halpnel3} correctly
describes the melting transition in $2D$. In the framework of the
KTHNY theory the two-dimensional melting occurs in the way which
is fundamentally different from the melting transition of
three-dimensional systems. In $2D$, the bound dislocation pairs
dissociate at some temperature $T_m$ transforming the quasi-long
range translational order in the short-range, and long-range
orientational order into the quasi-long range order. The new phase
with the quasi-long range orientational order is called the
hexatic phase. After consequent dissociation of the disclination
pairs at some temperature $T_i$ the system transforms into the
isotropic liquid. Both transitions are continuous, in contrast
with the conventional first order three dimensional melting.

The unambiguous confirmations of the KTHNY theory have been
obtained, for example,  from the recent experiments on the
colloidal model system with repulsive magnetic dipole-dipole
interaction \cite{keim1,zanh,keim2,keim3,keim4}. On the other
hand, the first-order melting in 2D is also possible
\cite{chui83,klein2,rto1,rto2,RT1,RT2,ryzhovTMP,ryzhovJETP}. In
Refs. \cite{ryzhovTMP,ryzhovJETP} it was shown that at low
disclination core energy system can melt through one first-order
transition as a result of the dissociation of the disclination
quadrupoles.

KTHNY theory is independent on the pair potential of the system
and seems universal, however, numerous experimental and simulation
studies demonstrate the controversial results: the systems with
very short range or hard core potentials melt through weak
first-order transition, while the melting scenarios for the soft
repulsive particles favor the KTHNY theory
\cite{rice,prest2,DF,LL,prest1,rice1,rice2,dfrt1,dfrt2,dfrt3,strandburg92,binderPRB,mak,jaster2,jaster3,andersen,andersen1,hfo1,hfo2,binder,ss1,ss2,ss3,LJ,gribova,lozfar1,lozfar2}.

In our previous publications the density functional approach for
the description of the $2D$ melting was proposed \cite{rto1,rto2}
and it was shown that the hard disk system melts through the first
order phase transition, while in the $2D$ Coulomb system the
melting transition occurs in accordance with the KTHNY scenario.
In Refs. \cite{RT1,RT2} the density functional calculations were
used for the description of the melting transition in the $2D$
square-well system. It was shown that this system can demonstrate
both first-order and continuous melting transitions depending on
the width of the attractive well.

In the present paper we extend our previous results to the melting
transition in the $2D$ square-shoulder system in order to study
the influence of the width of the shoulder on the phase diagram.
The potential is given by the equation:
\begin{equation}
U(r)= \left\{
\begin{array}{ll}
\infty,&r\leq d\\
\varepsilon,&d<r\leq \sigma\\
0,&r>\sigma
\end{array}\right..
\label{1}
\end{equation}
where $d$ is the diameter of the hard core, $\sigma$ is the width
of the repulsive step,  and  $\varepsilon$ its height. As it was
discussed before \cite{stishov,jcp2008}, in the low-temperature
limit $\tilde{T}\equiv k_BT/\varepsilon<<1$ the system reduces to
a hard-disk system with hard-disk diameter $\sigma$. At the same
time, in the limit $\tilde{T}>>1$ the system reduces to a
hard-disk model with a smaller hard-disk diameter $d$. In this
case, melting at high and low temperatures are described by the
simple hard-disk melting curve $P=cT/\sigma'^2$, where $\sigma'$
is the hard-disk diameters ($\sigma$ and $d$, respectively). A
crossover from the low-$T$ to high-$T$ melting behavior takes
place for $\tilde{T} ={\mathcal O}(1)$. The precise form of the
phase diagram depends on the ratio $s\equiv \sigma/d$. For large
enough values of $s$ one should expect to obtain the melting curve
with a maximum that should disappear as $s\rightarrow 1$
\cite{stishov,jcp2008}. In what follows we will use the reduced
units $r'=r/d, \rho'=\rho/d^2$ and omit tilde.

The different smoothed versions of the potential (\ref{1}) (core
softened potentials) were discussed recently in order to study the
water-like anomalies which appear due to the existence of two
length scales in the potential
\cite{jcp2008,buld2009,fr1,wepre,we_inv,we2011,RCR,we2013-2}.

In two dimensions the melting scenarios of the systems with the
core softened potentials were studied in Refs.
\cite{dfrt1,dfrt2,dfrt3} in the framework of computer simulations.
It was shown, that at low width of the soft core, when the system
behaves like an ordinary soft disk system, melting occurs through
one weak first order transition. However, with increasing the
width of the repulsive shoulder, the phase diagram becomes much
more complex. As in Ref. \cite{prest2}, we found that the phase
diagram consists of three different crystal phases, one of them
with square symmetry and the other two triangular. At low
densities, when the soft core of the potential is effective,
melting of the triangular phase is a continuous two-stage
transition, with an intermediate hexatic phase, in accordance with
the KTHNY scenario for this melting transition. At high density
part of the phase diagram one finds the square and triangular
phases, which melt through one first-order transition. The
thermodynamic and dynamic anomalies do exist in this case,
however, the order of this anomalies is inverted in comparison
with the three-dimensional case \cite{dfrt3}.

At the same time, in Ref. \cite{buld2d} the phase diagram of a
square-shoulder square-well potential was studied in two
dimensions. It has been previously shown that this potential
exhibits liquid anomalies consistent with a metastable
liquid-liquid critical point \cite{scala}. It was shown that the
melting occurs through the first order transition, despite a small
range of metastability.

In order to continue, let us for completeness briefly recall the
main ideas of the $2D$ density functional theory of freezing
\cite{rto1,rto2,RT1,RT2}. As it was mentioned above, in $2D$ the
long range translational order can not exist due to the thermal
fluctuations. Therefore, at low temperatures the local density of
a solid, which is proportional to the one-particle distribution
function, can be expanded in a Fourier series in reciprocal
lattice vectors {\bf G}:
\begin{equation}
\rho({\bf r}) = \sum_{\bf G} \rho_{\bf G}({\bf r}) e^{i{\bf G r}},
\label{2}
\end{equation}
where $\rho_{\bf G}({\bf r})$ are the order parameters for the
liquid-solid phase transition. Because of the thermal
fluctuations, the order parameters $\rho_{\bf G}({\bf r})$ slowly
vary at distances of order $G^{-1}$ and have the amplitude and the
phase:
\begin{equation}
\rho_{\bf G}({\bf r}) = \rho_{\bf G} e^{i{\bf G u}({\bf
r})}.\label{3}
\end{equation}
Here $u({\bf r})$ has the meaning of the displacement field,
which, in general, can be decomposed into the smooth part
corresponding to the phonon field, and singular part, which can be
interpreted as the Kosterlitz-Thouless vortices \cite{kosthoul73}
or dislocations.

Taking into account the long range fluctuations, one can write the
Landau expansion in the form:
\begin{eqnarray}
\Delta F&=&\frac{1}{2}\int\,d^2r\,\sum_{\bf G} \left[A|{\bf
G}\times \nabla \rho_{\bf G}|^2+ B|{\bf G}\cdot \nabla \rho_{\bf
G}|^2 + \right.\nonumber\\
&+&\left.|\rho_{\bf G}({\bf G}\cdot\nabla)
\rho_{\bf G}|\right]+\nonumber\\
&+&\frac{1}{2} a_T\sum_{\bf G}|\rho_{\bf G}|^2 + b_T\sum_{{\bf
G}_1+{\bf G}_2 +{\bf G}_3=0} \,\rho_{{\bf G}_1}\rho_{{\bf
G}_2}\rho_{{\bf G}_3}+\nonumber\\
&+&O(\rho^4) \label{4}.
\end{eqnarray}
$\Delta F$ corresponds to the difference of the free energy of
crystal and isotropic liquid. The first term in the expansion
(\ref{4}) has the form of the free energy of a deformed solid. The
Lame coefficient $\mu$ is a function of the parameters $A$, $B$,
and $C$ and is proportional to the squared modulus of the order
parameter (\ref{2}).

With the help of equation (\ref{4}), the $2D$ melting scenario can
be described in the following way. First of all, one can neglect
the fluctuations of the order parameter (like in three
dimensions). In this case, from Eq. (\ref{4}) one can see that
there is a possibility of the ordinary transition when at some
temperature $T_{MF}$ the modulus of the order parameter becomes
zero. Because of the third-order term in the Landau expansion
(\ref{4}), the transition is of the first order. However, there is
another possibility: at temperature $T_m$ the singular
fluctuations of the phase of the order parameter (vortices), which
corresponds to the free dislocations, appear in accordance with
the standard Kostelitz-Thouless paradigm. In this case the modulus
of the order parameter is not zero, however, the system will
respond to shear stress with no resistance ($\mu=0$), and it is
therefore possible to call the phase above $T_m$ as a liquid.
Really this is a hexatic phase with the quasi-long range
orientational order \cite{halpnel1,halpnel2,halpnel3}. There are
two possibilities: (i) $T_m<T_{MF}$, and the system melts through
the continuous Kosterlitz-Thouless transition; (ii) $T_m>T_{MF}$,
the system melts through the first-order transition.

$T_m$ and $T_{MF}$ can be calculated using the microscopic
expressions for the (Helmholtz) free energy $F$ of the solid and
for the elastic moduli. These expressions may be obtained in the
framework of the density functional theory of freezing
\cite{rto1,rto2,RT1,RT2}. In this case the local density of the
solid, $\rho({\bf r})$, can be represented as localized Gaussians
(of width $1/\alpha^{1/2}$) at lattice sites, i.e., with Fourier
components $\rho_{\bf G}= \rho\exp(-G^2/4\alpha$), where
$\rho=\int d{\bf r} \rho({\bf r})/V$ is the average density.
Because our main purpose is to obtain a qualitative description of
the melting transition in the system with the repulsive step
potential, we use the simplest but correct enough version of the
density functional theory \cite{rto1,rto2,RT1,RT2,baus,singh}:
\begin{eqnarray}
&&\beta \Delta F=\int\,d{\bf r}
\rho({\bf r})\,\ln[\rho({\bf r})/\rho]- \\
&&-\frac{1}{2} \int \,d{\bf r}\,\int \,d{\bf r}'\, c^{(2)}(|{\bf
r}-{\bf r}'|, \rho) [\rho({\bf r})-\rho][\rho({\bf
r}')-\rho],\nonumber \label{5}
\end{eqnarray}
where $c^{(2)}(|{\bf r}-{\bf r}'|, \rho)$ is the direct
correlation function \cite{book}, $\beta=1/k_B T$. The
localization parameter $\alpha$ is fixed by minimizing the total
free energy with respect to it. In principle, the parameters of
the first order melting transition should be determined from Eq.
(\ref{5}) using the double tangent construction, however, for the
approximate qualitative description of the phase diagram it seems
sufficient to apply the equation $\Delta F=0$ in order to
determine $T_{MF}$.

The temperature $T_m$ can be determined from the KTHNY criterion
\cite{halpnel1,halpnel2,halpnel3} which determines the instability
of the crystal lattice with respect to the appearance of free
dislocations:
\begin{equation}
K(T_{ m})=\frac{a_0^2}{k_{ B} T_{ m}} \frac{4 \mu(T_{ m}) (\mu(T_{
m})+\lambda(T_{ m}))} {2 \mu(T_{ m})+\lambda(T_{ m})} = 16 \pi
\label{8},
\end{equation}
where $\mu(T)$ and $\lambda(T)$ are the Lame coefficients and
$a_0$ is the lattice constant for the triangular lattice:
$a_0^2=2/(\sqrt{3}\rho)$.

Expressions for the Lame coefficients have the form
\cite{RT1,RT2}:
\begin{eqnarray}
\mu &=& \frac{k_{ B} T}{16 \rho} \sum_G \, \rho_G^2 m_G G^2
(\gamma_G+ 2\delta_G)
\label{9},\\
\lambda &=& \frac{k_{ B} T}{16 \rho} \sum_G \, \rho_G^2 m_G G^2
(\gamma_G- 6 \delta_G)+ \nonumber\\
&+& k_BT\rho(1-\rho
\tilde{c}^{(2)}(0)) \label{10},
\end{eqnarray}
where
\begin{eqnarray}
\gamma_G =&& 2 \pi \rho \int \,r^3d r\, c^{(2)}(r;\rho) J_0(Gr)
,\nonumber\\
\delta_G =&& 2 \pi \rho \int \,r^3d r\, c^{(2)}(r;\rho)
J_1(Gr)/(Gr),\nonumber
\end{eqnarray}
and $J_0(x)$ and $J_1(x)$ are the Bessel functions, $m_G$ is the
number of reciprocal vectors with the same length, and
$\tilde{c}^{(2)}(q)$ is the Fourier transform of the direct
correlation function.

\begin{figure}

\includegraphics[width=8cm]{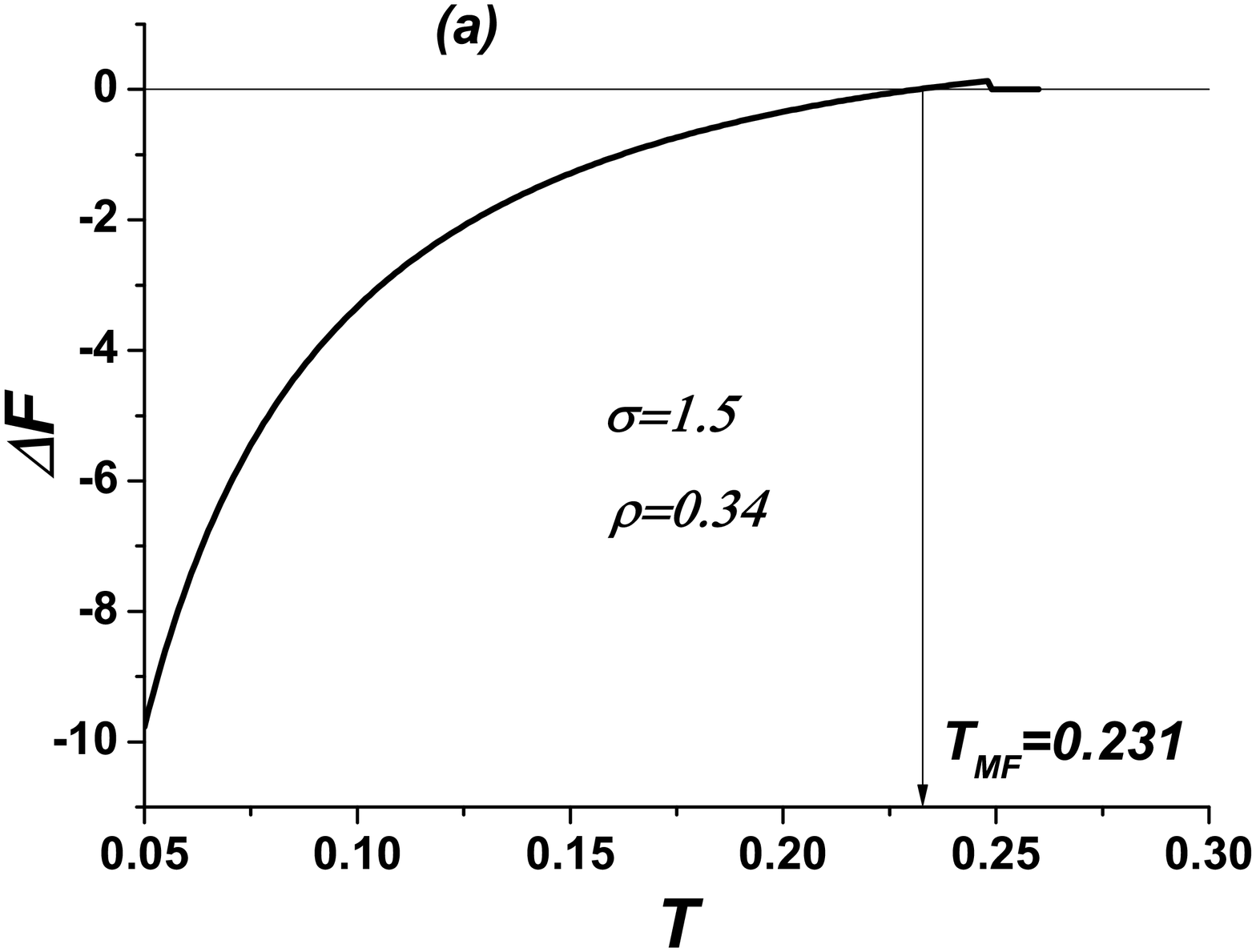}%

\includegraphics[width=8cm]{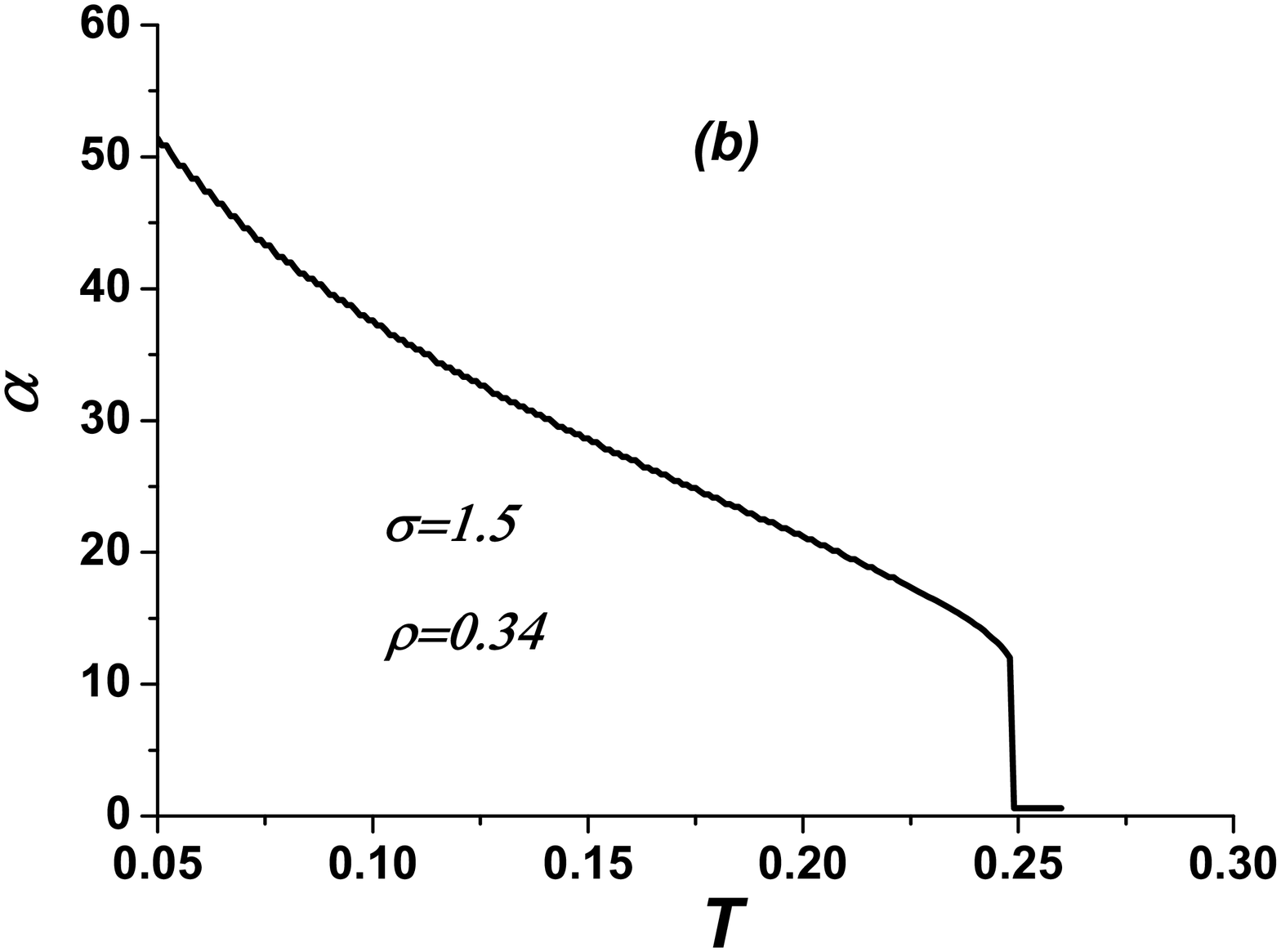}%

\includegraphics[width=8cm]{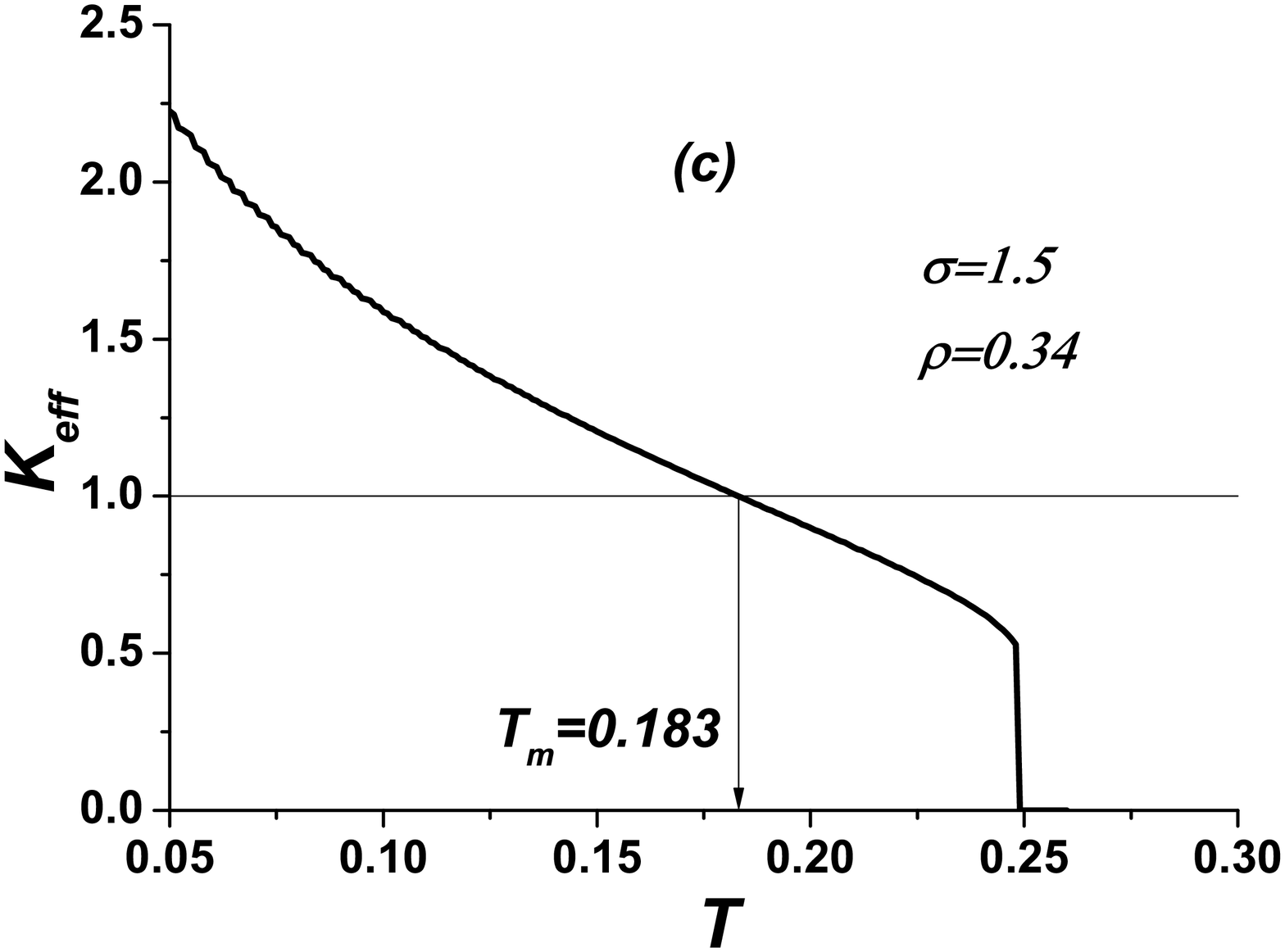}%

\caption{\label{fig:fig1} $\Delta F$ (a), $\alpha$ (b), and
$K_{eff}=K/16\pi$ (c) as functions of temperature for $\rho=0.34$.
}
\end{figure}

\begin{figure}

\includegraphics[width=8cm]{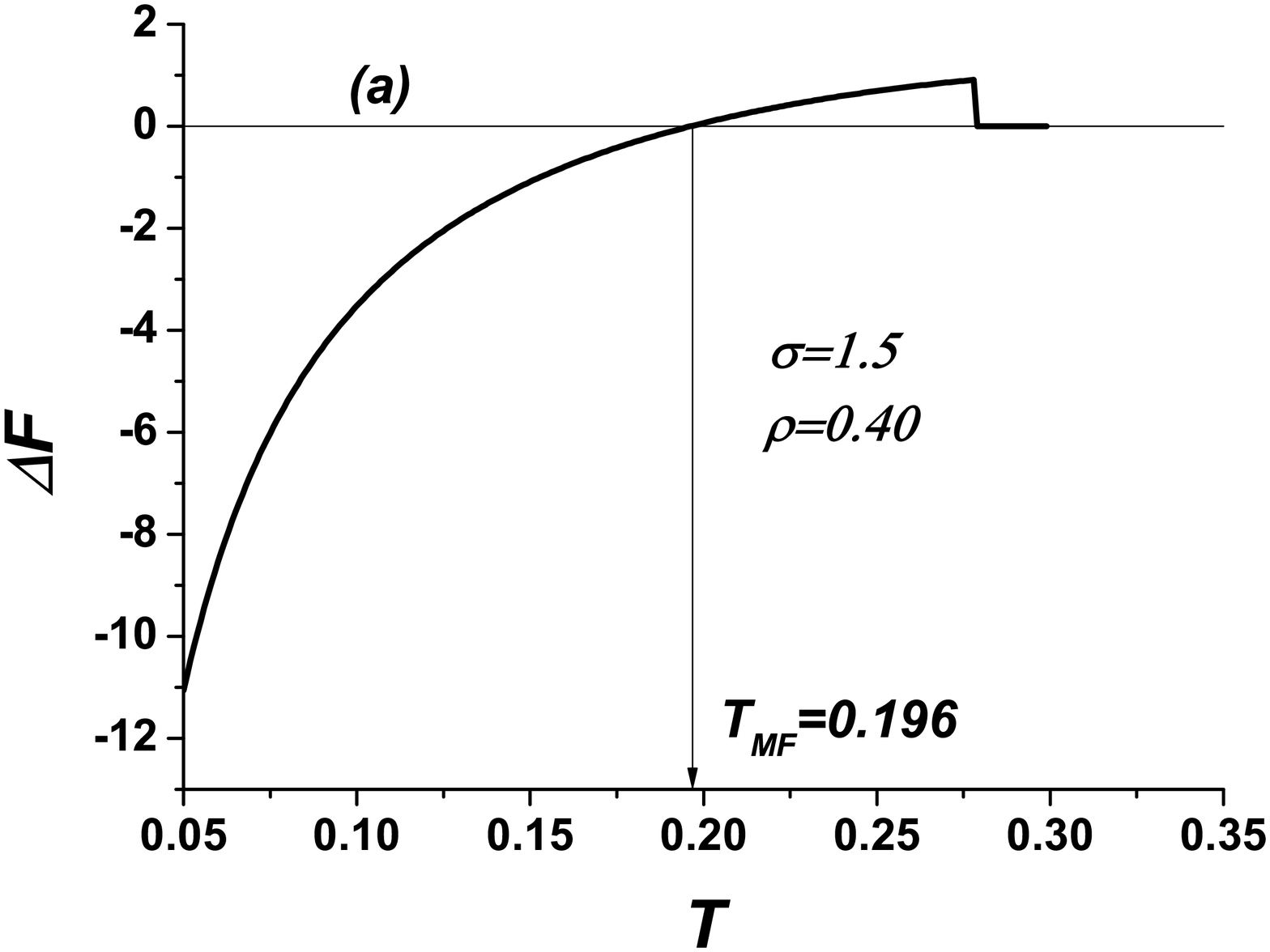}%

\includegraphics[width=8cm]{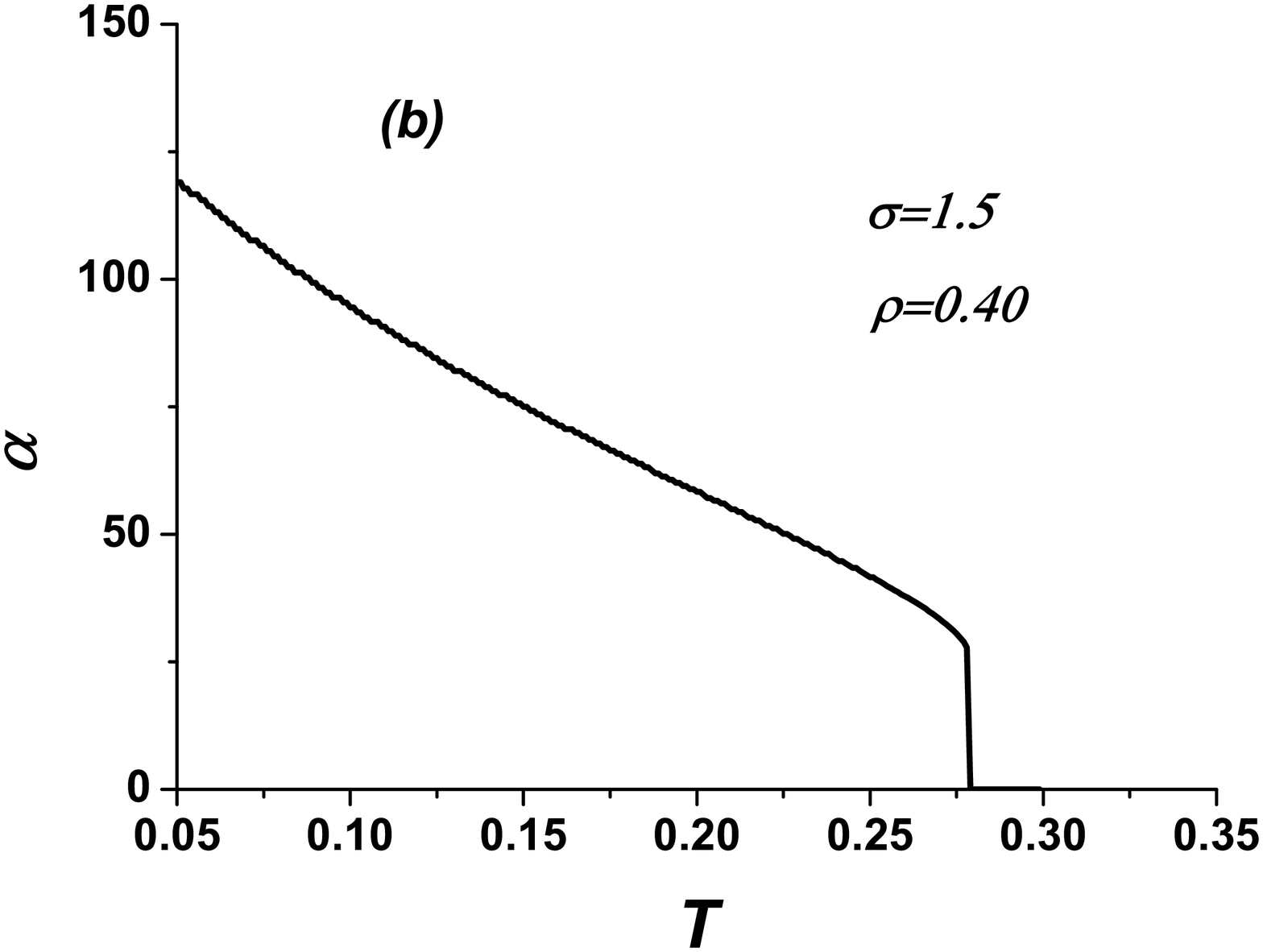}%

\includegraphics[width=8cm]{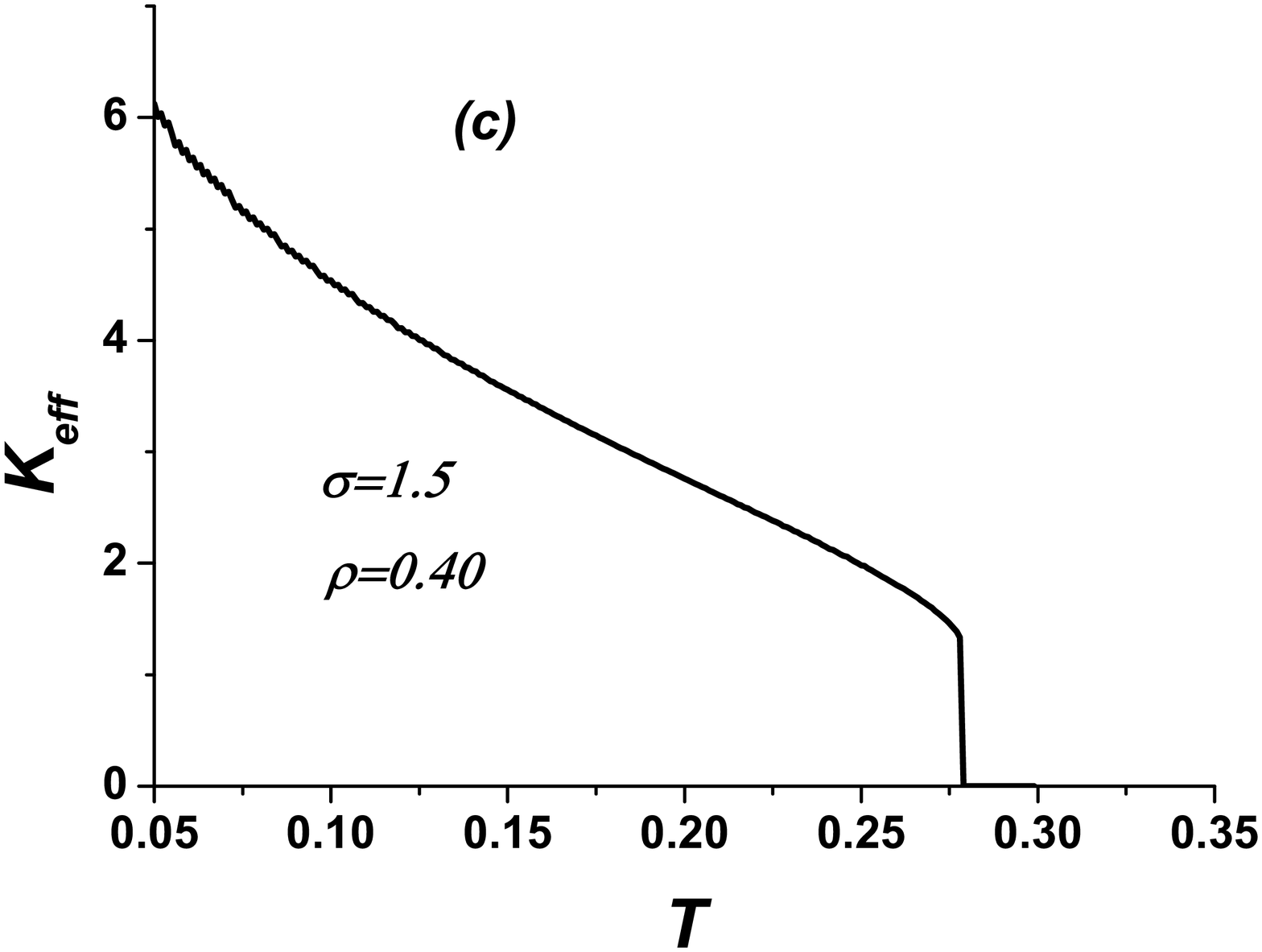}%

\caption{\label{fig:fig2} $\Delta F$ (a), $\alpha$ (b), and
$K_{eff}=K/16\pi$ (c) as functions of temperature for
$\rho=0.40$.}
\end{figure}

\begin{figure}

\includegraphics[width=8cm]{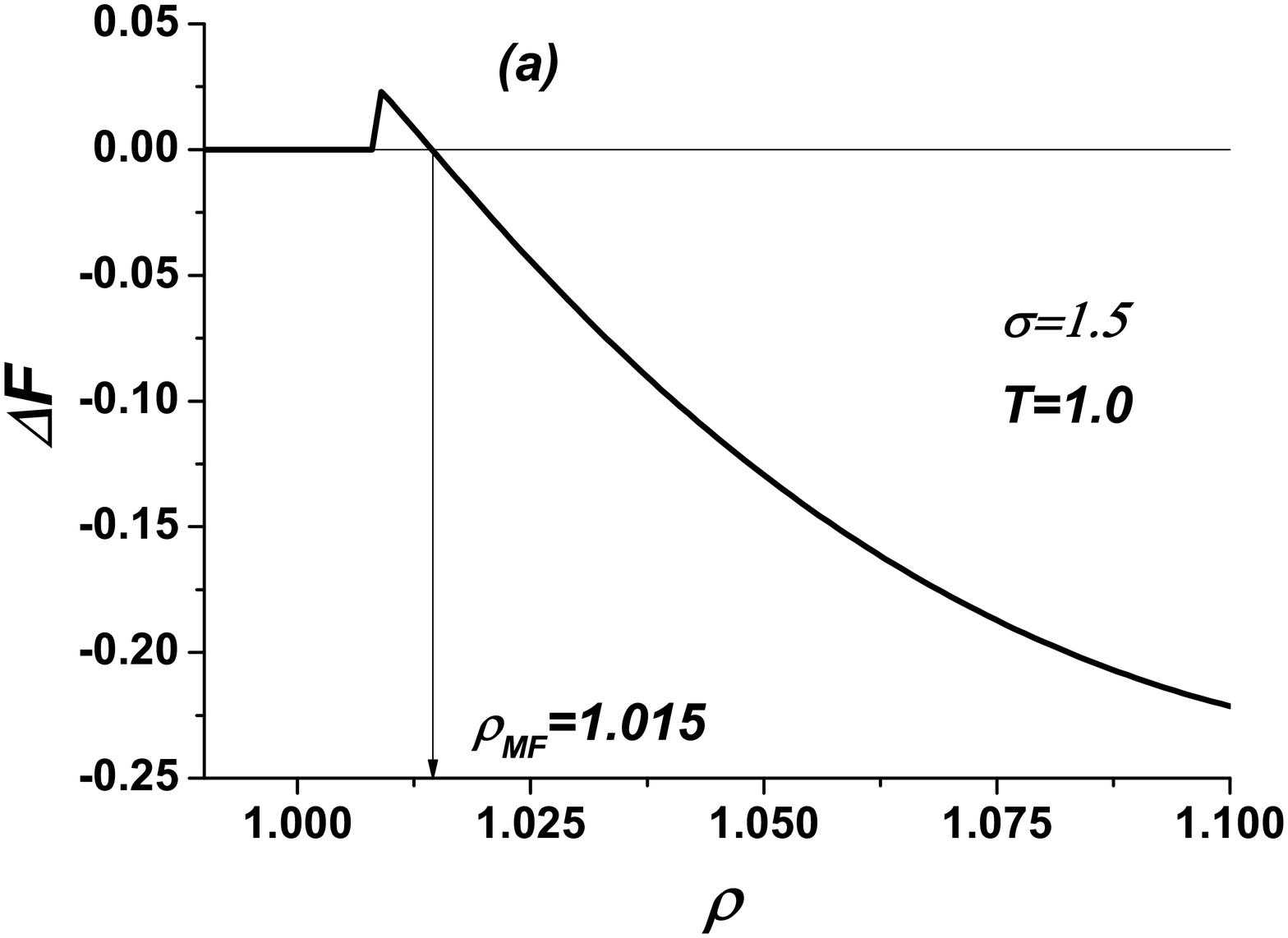}%

\includegraphics[width=8cm]{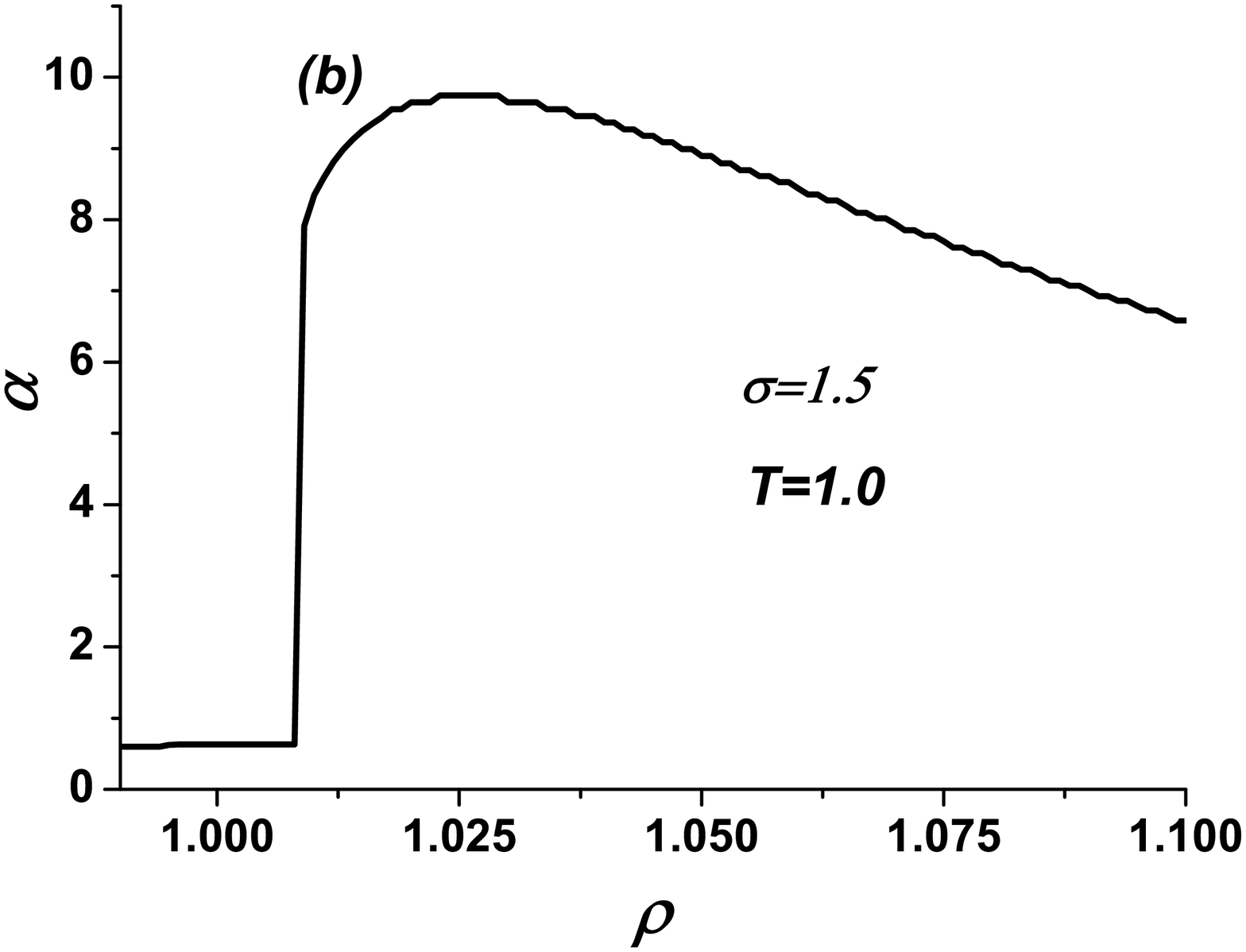}%

\includegraphics[width=8cm]{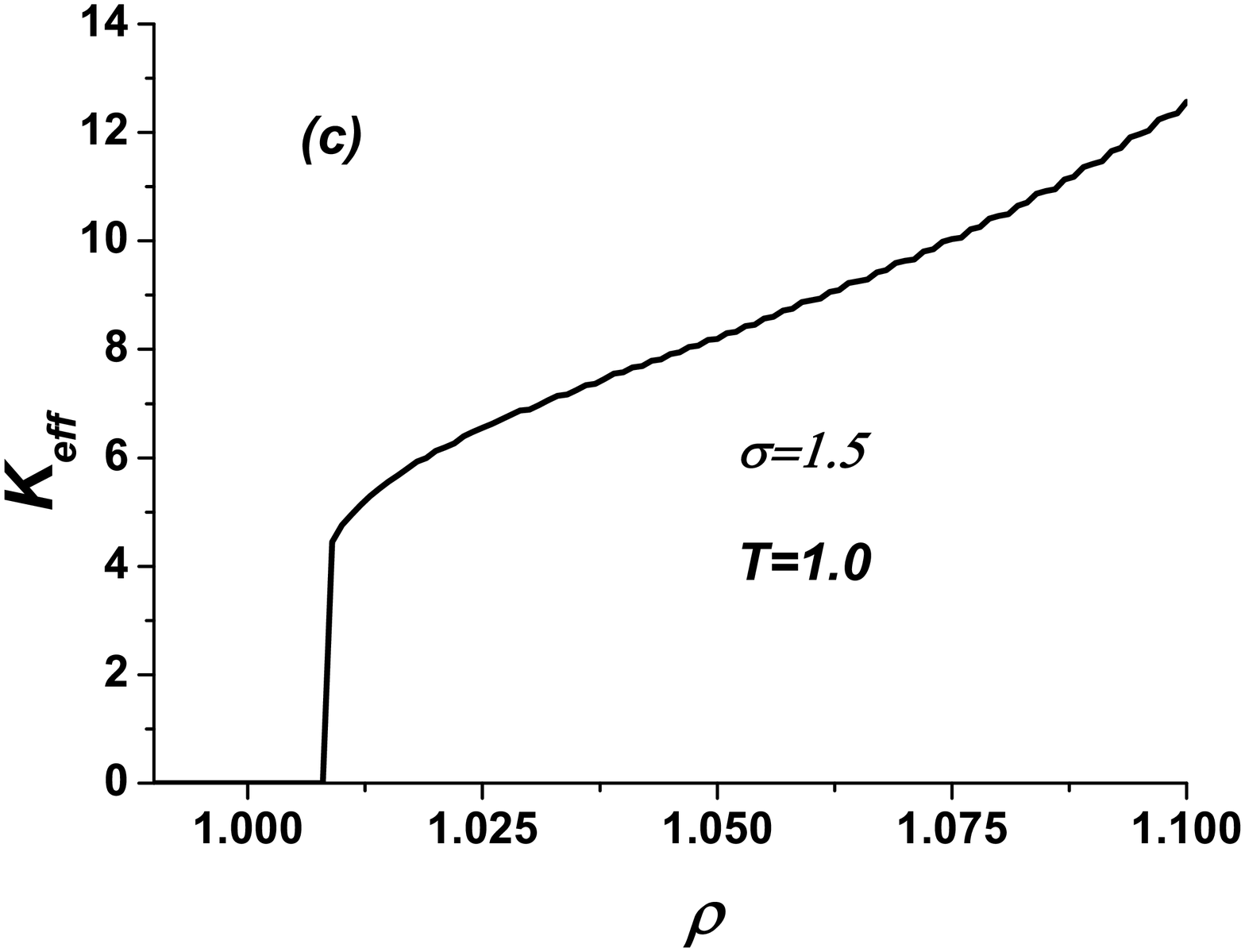}%

\caption{\label{fig:fig3} $\Delta F$ (a), $\alpha$ (b), and
$K_{eff}=K/16\pi$ (c) as functions of density for $T=1.0$.}
\end{figure}

For the further calculations one needs an approximate expression
for the direct correlation function. We use the simple
approximation for the direct correlation function \cite{book} of
the hard-core system, suggested by Lovett \cite{lovett} (see also
\cite{RT1,RT2,TMF,pre2014}):
\begin{equation}
c^{(2)}(r,\rho)=\left\{
\begin{array}{ll}
c^{(2)}_{HD}(r,\rho) , & r\leq d \\
-\frac{\phi(r)}{k_BT}, & r>d
\end{array}%
\right.,   \label{111}
\end{equation}
where $c^{(2)}_{HD}(r,\rho)$ is the hard disks direct correlation
function and $\phi(r)$ is the repulsive shoulder or attractive
part of the potential. This approximation should be a good one
when $-\frac{\phi(r)}{k_BT}$ is small. The approximation, though
rough, is similar in spirit to the mean spherical model
approximation which has been found to be a good approximation in
many cases \cite{book}. In the case of the potential (\ref{1}),
Eq. (\ref{111}) takes the form:
\begin{equation}
c^{(2)}(r,\rho) \approx \left\{
\begin{array}{ll}
c_{HD}^{(2)}(r,\rho),&r\leq d\\
\frac{-\varepsilon}{k_BT},&d<r\leq h\\
0,&r>h
\end{array}\right.
\label{11}.
\end{equation}

For $c_{HD}^{(2)}(r,\rho)$ we use the approximate analytic
equation obtained in Refs. \cite{c1,c2}:
\begin{eqnarray}
&&c^{(2)}_{ HD}(x;\eta)= - \left[\frac{\partial}{\partial \eta}
(\eta Z(\eta)) \right] \Theta (1-x)\times\nonumber\\
&&\times\left\{1-a^2\eta+\frac{2}{\pi}a^2 \eta \left[
\arccos\frac{x}{a}-\frac{x}{a}\left(1-\frac{x^2}{a^2}\right)^{1/2}
\right]\right\}, \nonumber\\
&&Z(\eta)= (1+c_2 \eta^2)/(1-\eta)^2, \nonumber\\
&&a= (2+\eta \alpha_2(\eta))/(1+\eta+\eta \alpha_2(\eta)),
\label{12}
\end{eqnarray}
where $c_2=0.128; \alpha_2(\eta)=-0.2836+0.2733 \eta; \eta=\pi
\rho d^2/4.$

\begin{figure}

\includegraphics[width=8cm]{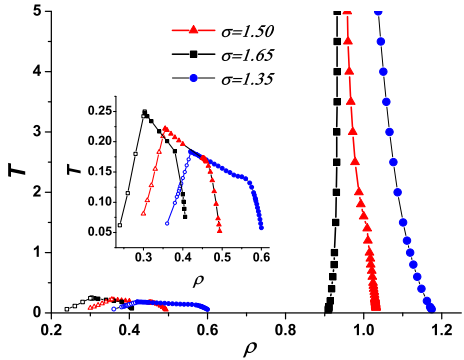}%

\caption{\label{fig:fig4} Phase diagram of the system with the
potential (\ref{1}) for three repulsive shoulder widths:
$\sigma=1.50$ (triangles), $\sigma=1.35$ (circles), $\sigma=1.65$
(squares). Open symbols correspond to the continuous transitions,
while the bold symbols mark the first order transitions.}
\end{figure}

In Figs.~\ref{fig:fig1} and ~\ref{fig:fig2} we represent the
behavior of $\Delta F$ (see Eq. (\ref{4})), localization parameter
$\alpha$, and $K_{eff}=K(T)/16\pi$ (Eq. (\ref{8})) as a function
of $T$ for $\sigma=1.5$ and $\rho=0.34$ (Fig.~\ref{fig:fig1}) and
$\rho=0.4$ (Fig.~\ref{fig:fig2}). One can see, that for
$\rho=0.34$ the solution of equation $\Delta F=0$ which determines
the first-order transition temperature $T_{MF}$, is $T_{MF}=0.231$
(Fig.~\ref{fig:fig1}(a)), while the solution of the equation
$K_{eff}=1$ (Fig.~\ref{fig:fig1}(c)) $T_m=0.183<T_{MF}$. As it was
discussed above, in this case the melting should occur in
accordance with the KTHNY scenario. From Fig.~\ref{fig:fig1}(b)
one can conclude that the localization parameter $\alpha$ is well
defined till the limit of metastability of the crystal lattice
$T_{met}=0.248$. On the other hand, for $\rho=0.40$ the situation
is different. In this case $T_{MF}=0.196$  while the equation
$K_{eff}=1$ does not have solution till the limit of metastability
$T_{met}=0.278$. In this case melting occurs through the first
order transition.

At high densities there is only weak dependence of the melting
density on temperature (see below). In this case, in calculations
it is more convenient to fix the temperature and change the
density. The typical results are shown in Fig.~{\ref{fig:fig3} for
$\sigma=1.5$ and $T=1.0$. One can see that at high densities for
$T=1.0$ there is the first order melting transition at
$\rho_{MF}=1.015$. The equation $K_{eff}=1$ does not have solution
till the limit of metastability $\rho_{met}=1.009$.

In Fig.~\ref{fig:fig4} we present the resulting phase diagram for
three widths of the repulsive shoulder in the potential (\ref{1}):
$\sigma=1.50$, $\sigma=1.35$, and $\sigma=1.65$. In accordance
with the qualitative discussion after Eq. (\ref{1}), the phase
diagram consists in two parts - the low density triangle lattice
with the maximum on the melting curve and the high density
triangle lattice. It is interesting that at lowest density part of
the phase diagram the KTHNY scenario takes place, while with the
increasing density the melting becomes the first order transition.
Taking into account the fact, that with increasing the density the
hard core of the potential (\ref{1}) becomes effective, one can
conclude that this result is consistent with the mentioned above
possibility that the systems with soft potentials probably melt in
accordance with the KTHNY scenario, while the hard core systems
melt through the first order transition. It should be noted that
with increasing temperature the system has to behave more and more
closely to the hard disk system. There are different estimates for
the melting density of hard disk systems
\cite{rto1,rto2,binderPRB,mak,jaster3} which vary from
$\rho=0.905$ to $\rho=0.933$. One can see that despite very simple
approximations, the results presented in Fig.~\ref{fig:fig4} for
high temperatures, are in good enough qualitative agreement with
previous estimates.

We also considered the possible square lattice but found that it
is less stable than the triangle lattice for all densities.

It should be noted that the similar behavior was found in computer
simulation of the smoothed version of the potential (\ref{1})
\cite{dfrt1,dfrt2,dfrt3}. As it was mentioned in the introduction,
in Refs. \cite{dfrt1,dfrt2,dfrt3} it was shown that at low
densities melting occurs through two continuous transitions with
the intermediate hexatic phase, however, at high densities only
first order transition takes place. As in the present case, the
"gap" between the two parts of the phase diagram takes place which
increases with increasing the width of the repulsive shoulder. In
principle, some crystal lattice can exists in this range of
densities at low enough temperatures, however, we could not find
it in the present work. Another open question exists for further
study. It is related with the crossover from the continuous to
first order transition in the phase diagram in
Fig.~\ref{fig:fig4}. It must be the tricritical point on the
melting line, however, in the present study we could give only
rough enough estimate for the location of this point without
investigation of the properties of this point.

In conclusion, in the present study we consider the melting
transition of the repulsive shoulder potential system (\ref{1}).
In the framework of the density functional theory of freezing we
calculate the phase diagram and show that it consists of two parts
with different melting scenarios (see Fig.~\ref{fig:fig4}).  At
low densities the system melts through the continuous transition
in accordance with the Kosterlitz-Thouless-Halperin-Nelson-Young
(KTHNY) scenario, while at high densities the conventional first
order transition takes place. Taking into account the fact, that
with increasing the density the hard core of the potential
(\ref{1}) becomes effective, one can conclude that this result is
consistent with the possibility that the systems with soft
potentials probably melt in accordance with the KTHNY scenario,
while the hard core systems melt through the first order
transition.

\bigskip

The authors are grateful to S.M. Stishov and V.V. Brazhkin for
valuable discussions. The work was supported by the Russian
Science Foundation (Grant No 14-22-00820).

\end{document}